\documentclass[11pt,a4paper]{article}
\usepackage{amsmath}

\usepackage[psamsfonts]{amssymb}
\usepackage{amsmath}
\usepackage{epsfig}

\author{H. Mohseni Sadjadi\footnote{mohseni@phymail.ut.ac.ir}
\\ {\small Department of physics, University of Tehran ,}
\\ {\small P.O.B. 14395-547, Tehran 14399-55961, Iran}}
\title{Schwarzschild black hole and generalized second law
in phantom-dominated universe}

\begin{document}
\maketitle
\begin{abstract}
In this paper which is an extension of the work \cite{pavon}, we
study the conditions required for validity of the generalized
second law in phantom dominated universe in the presence of
Schwarzschild black hole. Our study is independent of the origin
of the phantom like behavior of the considered universe. We also
discuss the generalized second law in the neighborhood of
transition (from quintessence to phantom regime) time. We show
that even for a constant equation of state parameter, the
generalized second law may be satisfied provided that the
temperature is not taken as de Sitter temperature. It is shown
that in models with (only) a transition from quintessence to
phantom regime the generalized second law does not hold in the
transition epoch.
\end{abstract}
\section{Introduction}
Astrophysical data show that the universe is accelerating
\cite{acc}. Based on some data, it is possible to consider an
evolving equation of state parameter, $\omega$, less than $-1$ at
present time from $\omega>-1$ in the near past \cite{cross}. In
this view we may assume that the universe is filled with a perfect
fluid with a negative pressure and $\omega<-1$, dubbed as phantom
dark energy \cite{phant}. A candidate for phantom dark energy is a
phantom scalar field with wrong sign for kinetic energy term
\cite{phant scalar}. Another method to study the present inflation
is to use a running cosmological constant based on principles of
quantum field theory (specially on the renormalization group)
which can mimic the phantom like behavior of the universe
\cite{mimic}.

This description of the universe may contain finite time future
singularity accompanied with dark energy density singularity
called big rip. The big rip may be avoided by the effect of
gravitational backreactions which can end the phantom dominated
regime \cite{back}. We can consider horizons for the accelerating
universe and associate entropy (as a measure of our ignorance
about what is going behind it) and temperature to them \cite{dav1,
dav2, poll, pav1, moh, nojiri, horizon}. In this way one is able
to study the thermodynamics of a system consisting of dark energy
perfect fluid and the horizon.

In phantom dominated universe black holes lose their masses by
accreting phantom fluid \cite{babi}. Therefore their areas and
consequently their entropies will decrease. So it may be of
interest to know that if the generalized second law of
thermodynamics (GSL) is satisfied in this situation. Indeed if the
thermodynamics parameters assigned to the universe are the same as
the ordinary thermodynamics parameters known in physical systems,
then one expects that thermodynamics laws be also satisfied for
the universe.

Thermodynamics of an accelerating universe has been studied in
several papers \cite{thermo}. In \cite{dav1} and \cite{dav2}, the
generalized second law for cosmological models that depart
slightly from de Sitter space and also when the horizon shrinks,
was studied respectively.  The thermodynamics of super-accelerated
universe in a de Sitter and quasi de Sitter space-time was the
subject of the paper \cite{poll}.

In \cite{pav1}, it was shown that for a phantom dominated universe
with constant $\omega$ the total entropy is a constant and for
time dependent $\omega$, via two specific examples, the validity
of GSL was verified. In \cite{moh} the conditions of validity of
GSL in more general cases, including the transition epoch (from
quintessence to phantom), and for temperatures proportional to de
Sitter temperature were studied independently of the origin of
dark energy.

In a  recent paper the author of \cite{pavon}, using phantom
scalar field model, showed that GSL is violated in the presence of
a Schwarzschild black hole in the cases studied in \cite{pav1} and
in phantom dominated era . In this paper we try to study the same
problem but by considering a temperature other than de Sitter
temperature. Our study is independent of the origin of phantom
like behavior of the universe. We also consider the possibility of
transition from quintessence to phantom regime and discuss the
validity of GSL in the neighborhood of transition time in the
presence of the black hole.

We use the units $\hbar=c=G=k_{B}=1$.

\section{GSL in the phantom dominated FRW universe in the presence of a Schwarzschild black hole}
We consider spatially flat Friedman Robertson Walker (FRW) metric
with scale factor $a(t)$:
\begin{equation}\label{1}
ds^2=-dt^2+a^2(t)(dx^2+dy^2+dz^2).
\end{equation}
The Hubble parameter is given by $H=\dot{a}/a$. The over dot shows
derivative with respect to the comoving time $t$. The equation of
state of the universe which is assumed to behave as a perfect
fluid at large scale is given by
\begin{equation}\label{2}
p=\omega \rho,
\end{equation}
where $\omega$ is the equation of state parameter. For an
accelerating universe, i.e. $\ddot{a}>0$, we have $\omega<-1/3$.
The future event horizon, $R_h$, is given by
\begin{equation}\label{3}
R_h(t)=a(t)\int_t^{\infty}\frac{dt'}{a(t')},
\end{equation}
where $\lim _{t\rightarrow \infty}a(t)=\infty$ and $\int_t^\infty
dt'/a(t')<0$. In the presence of big rip singularity at $t_s$, we
must replace $\infty$ by $t_s$ in the integration. For a de Sitter
space-time $a(t)\propto \exp(Ht)$, and the future event horizon
reduces to de Sitter horizon : $R_h=1/H$. In this space time the
equation of state parameter is $\omega=-1-2\dot{H}/(3H^2)=-1$.

If the system remains in quintessence phase, defined by
$-1<\omega<-1/3$ (or $\dot{H}<0$), the future event horizon
satisfies $\dot{R_h}\geq 0$.  Instead, for an universe which will
remain in phantom dominated era, defined by $\omega<-1$ (or
$\dot{H}>0$), we have $\dot{R_h}\leq 0$. It is worth to note that
these behaviors of the future event horizon depend on the entire
future circumstances, e.g., if the phantom  ends to quintessence
phase, we may have $\dot{R_h}\geq 0$ even in the phantom dominated
era.

One can consider an entropy for the future event horizon as a
measure of information hidden behind it:
\begin{equation}\label{4}
S_h=\pi R_h^2.
\end{equation}
By adopting this point of view, we obtain the total entropy of the
universe, $S$,  as the sum of the entropy inside the horizon,
$S_{in}$, and $S_h$:
\begin{equation}\label{5}
S= S_{in}+S_h
\end{equation}

The perfect fluid is supposed to be in thermal equilibrium with
the future event horizon (note that FRW model requires thermal
equilibrium). When the space-time (\ref{1})is de Sitter, i.e. when
the future event horizon is the same as de Sitter horizon, we can
consider the temperature as $T=H/(2\pi)$. Note that this has been
only verified for de Sitter horizons \cite{gibb}. For a non-de
Sitter space-time, i.e. when $R_h\neq 1/H$, we {\it assume} that
the future event horizon temperature is proportional to de Sitter
temperature (which is the only temperature scale we have at our
disposal) \cite{dav1}
\begin{equation}\label{6}
T=\frac{bH}{2\pi},
\end{equation}
where $b$ is a real constant.

Besides the dark energy and dark matter, we introduce a
Schwarzschild black hole inside the future event horizon. The mass
of the black hole, $M$, is assumed to be enough small so that the
metric (\ref{1}) remains unchanged. Using $\rho=3H^2/(8\pi)$,
where $\rho$ is the energy density inside the future event
horizon, this condition reduces to
\begin{equation}\label{7}
MH\ll \frac{R_h^3H^3}{2}
\end{equation}

$S_{in}$ may be divided into two parts: entropy of the black hole,
denoted by $S_{bl}$ and the entropy of perfect fluids denoted by
$S_d$
\begin{equation}\label{8}
S_{in}=S_{bl}+S_d.
\end{equation}
In a fluid with the energy density $\rho$ and the pressure $P$,
the change rate of the black hole mass is \cite{babi}
\begin{eqnarray}\label{9}
\dot{M}&=&4\pi A r_h^2(P+\rho)\nonumber \\
&=&-4A M^2\dot{H},
\end{eqnarray}
where $r_h$ is the radius of the black hole horizon and $A$ is a
positive numerical constant. So, in terms of the Hubble parameter,
the black hole mass may be obtained as
\begin{equation}\label{10}
M=\frac{1}{C+4AH}
\end{equation}
where $C$ is a numerical constant.

The entropy of the black hole is $S_{bl}=4\pi M^2$ \cite{bek},
therefore
\begin{equation}\label{11}
\dot{S_{bl}}=-32\pi AM^3\dot{H}.
\end{equation}

The entropy of the phantom fluid inside the cosmological horizon
is related to the energy and the pressure via the first law of
thermodynamics
\begin{equation}\label{12}
TdS_{d}=dE+PdV=(P+\rho)dV+Vd\rho,
\end{equation}
where $V=(4/3)\pi R_h^3$ is the volume inside the future event
horizon. Using (\ref{12}) we obtain \cite{moh}
\begin{equation}\label{13}
T\dot{S_d}=\dot{H}R_h^2.
\end{equation}
Note that if $T>0$ then $\dot{S_d}>0$. The generalized second law
asserts that the sum of the ordinary entropy, the future event
horizon entropy and the black hole entropy cannot decrease with
time: $\dot{S_d}+\dot{S_{bl}}+\dot{S_h}\geq 0$. This leads to
\begin{equation}\label{new1}
\dot{H}(\frac{R_h^2}{T}-32\pi AM^3)+2\pi \dot{R_h}R_h\geq 0.
\end{equation}
Note that for a de Sitter space-time, GSL is satisfied:
$\dot{H}=0$ and $\dot{S_d}+\dot{S_{bl}}+\dot{S_h}= 0$.

In phantom era $\dot{H}>0$, therefore for a system remaining in
phantom phase, $T>0$ is a necessary condition for validity of GSL
( Note that in such system we have $\dot{R_h}\leq 0$). Also in the
presence of the black hole, GSL is violated in phantom models with
negative temperature. Using (\ref{new1}) we find
\begin{equation}\label{14}
\dot{H}\left(\frac{R_h^2}{H}-16bAM^3 \right)+bR_h\dot{R_h}\geq 0.
\end{equation}
This results in that in order that GSL holds at $t_0$, where
$\dot{H}(t_0)=0$, we must have $\dot{R_h}(t_0)\geq 0$. In the
phantom regime $\dot{H}>0$, hence $H$ is an increasing function of
time, so that we may write (\ref{14}) as
\begin{equation}\label{15}
\frac{b}{2}\frac{dR_h^2}{dH}+\frac{R_h^2}{H}\geq 16bAM^3
\end{equation}

To go further let us study the validity of GSL in some special
cases which are of interest: For example consider a phantom
dominated universe with a constant equation of state parameter,
$\omega(t)=\omega_0\neq -1$, with a big rip at $t=t_s$.  The
Hubble parameter is then
\begin{equation}\label{16}
H=\frac{2}{3(1+\omega_0)(t-t_s)}.
\end{equation}
Using
\begin{equation}\label{17}
\dot{R_h}=HR_h-1,
\end{equation}
we obtain
\begin{equation}\label{18}
R_h=3\frac{1+\omega_0}{1+3\omega_0}(t_s-t)
\end{equation}
which leads to
\begin{equation}\label{19}
HR_h=\beta,
\end{equation}
where $\beta$ is a constant, $\beta=-2/(3\omega_0+1)<1$, in
agreement with the expected decreasing behavior of the future
event horizon. Note that even for $\omega_0=-1$, which describes a
de Sitter space, (\ref{19}) is still valid. The solution of
(\ref{9}) is
\begin{equation}\label{20}
M=\frac{t_s-t}{-\frac{8A}{3(1+\omega_0)}+C(t_s-t)}.
\end{equation}
$C$ is given by:
\begin{equation}\label{21}
C=\frac{1}{M(t_i)}+\frac{8A}{3(1+\omega_0)}\frac{1}{t_s-t_i },
\end{equation}
where $t_i$ is an arbitrary time in phantom dominated era.
Combining (\ref{20}) and (\ref{16}) we arrive at
\begin{equation}\label{22}
MH=\frac{1}{4A-\frac{3C}{2}(1+\omega_0)(t_s-t)}.
\end{equation}

Using (\ref{19}) we can write (\ref{15}) in the form
\begin{equation}\label{23}
M^3H^3\leq \beta^2\frac{1-b}{16bA},
\end{equation}
which, for $b=1$,  does not hold and the generalized second law is
not respected, in agreement with the claim of \cite{pavon}. But it
seems that for $b<1$, GSL may be respected for suitably chosen
parameters, at least in the domain of validity of the
approximation (\ref{7}). To see this, we proceed as follows : For
$b<1$ and $\dot{S}>0$, in order to satisfy the GSL, we must have
\begin{equation}\label{24}
\frac{1}{4A-\frac{3C}{2}(1+\omega_0)(t_s-t)}<
(\beta^2\frac{1-b}{16bA})^{\frac{1}{3}}
\end{equation}
in addition, for validity of our approximation (\ref{7}), we
require
\begin{equation}\label{25}
\frac{1}{4A-\frac{3C}{2}(1+\omega_0)(t_s-t)}\ll
\frac{1}{2}\beta^3.
\end{equation}
Hence GSL is respected in times : $t$, satisfying (\ref{24}) and
(\ref{25}). Near $t=t_s$, the approximation (\ref{25}) is not
satisfied for $A\sim O(1)$. If $C>0$ and if GSL holds for a
specific $t=t_i$, it will hold for $t<t_i$.

For $b<1$ and $\dot{S}=0$ (corresponding to reversible adiabatic
expansion), we obtain
\begin{equation}\label{26}
MH= \gamma,
\end{equation}
where $\gamma^3=\beta^2(1-b)/(16bA)$. Now From (\ref{26}) and
(\ref{22}) we can determine $\gamma$ and $C$: $\gamma=1/(4A)$,
$C=0$. On the other hand the validity of the approximation
(\ref{7}) requires: $\gamma\ll \beta^3/2$, which is only valid for
large $A$.

As another example consider time depending $\omega(t)$ and
$\dot{S}=0$. In this case one can determine $R_h$ as a function of
time. Applying $\dot{S}=0$ in (\ref{15}) gives
\begin{equation}\label{27}
R_h^2(H)=H^{-\frac{2}{b}}\left(C_1+32A\int
M^3(H)H^{\frac{2}{b}}dH\right).
\end{equation}
$C_1$ is a numerical constant. Inserting (\ref{10}) into the above
integral yields
\begin{equation}\label{28}
R_h^2=C_1
H^{-\frac{2}{b}}+\frac{4}{(4AH+C)^2b}[2+\frac{8HA}{C}-b]+\frac{8}{C^2b}(1-\frac{2}{b})\Phi
(-\frac{4HA}{C},1,\frac{2}{b})
\end{equation}
where $\Phi$ is the Lerchphi function. But following the
approximation (\ref{7}), the solution (\ref{28}) is only valid
when $4A+C/H\gg 1$. For $A\sim O(1)$ and $C/H\gg 1$, by
considering the series representation of Lerchphi function, we
obtain
\begin{equation}\label{29}
R_h^2H^2=C_1
H^{2-\frac{2}{b}}+\frac{32Ab}{b+2}(\frac{H}{C})^3+O((\frac{H}{C})^4).
\end{equation}
This equation with (\ref{17}) determine $R_h$. Up to the order
$O((H/c)^3)$, by inserting (\ref{29}) into (\ref{17}) we find
\begin{equation}\label{30}
\dot{R_h}-C_1^{\frac{b}{2}}R_h^{1-b}+1=0.
\end{equation}
For $b=1$, the problem reduces to $\omega=\omega_0=C_1^{1/2}$,
discussed in the previous part. For $b\neq 1$, solution of
(\ref{30}) satisfies
\begin{equation}\label{31}
\frac{R_h\Phi(C_1^{\frac{b}{2}}R_h^{1-b},1,\frac{1}{1-b})}{1-b}=d-t.
\end{equation}
At $t=d$, we have $R_h=0$. Note that, in this approximation
$R_h^bH=C_1^{\frac{b}{2}}$. Comparing of this result with that
obtained in \cite{moh} indicates that the presence of the black
hole in the domain of validity of GSL and the approximation
(\ref{7}), up to the order $M^3$, does not change the behavior of
$R_h$.

 \subsection{GSL near the transition time}

Based on astrophysical data, which seem to favor an evolving dark
energy with $\omega$ less than $-1$ at present epoch from
$\omega>-1$ in the near past \cite{cross}, it may be interesting
to study the validity of GSL near the transition time(time of
$\omega=-1$ crossing). In the phantom regime $\dot{H}>0$ and in
the quintessence regime we have $\dot{H}<0$, therefore if the
Hubble parameter has a Taylor series at transition time, which is
taken to be at $t=0$, $\dot{H}(0)=0$ and we can write
\begin{equation}\label{32}
H=h_0+h_1t^a,
\end{equation}
where $h_0=H(t=0)$ and $a$, a positive even integer number, is the
order of the first nonzero derivative of $H$ at $t=0$.
$h_1=H^{(a)}/(a!)$ and $H^{(a)}=d^aH/dt^a$. In the case of
transition from quintessence to phantom phase we must have
$h_1>0$. Using (\ref{17}) it can be shown that $R(t)$ has the
following expansions:
\begin{equation}\label{33}
R_h(t)=R_h(0)+(h_0R_h(0)-1)t+O(t^2),
\end{equation}
for $\dot{R_h}(0)\neq 0$, and
\begin{equation}\label{34}
R_h(t)=R_h(0)(1+\frac{h_1}{a+1}t^{a+1})+O(t^{a+2}),
\end{equation}
for $\dot{R_h}(0)= 0$, at $t=0$. Near the transition time
(\ref{7}) reduces to $h_0^2R_h^3(0)\gg 2M(0)$.

The condition of validity of GSL near the transition time, $t=0$,
for $\dot{R_h}(0)=0$, can be investigated by inserting
$H=h_0+h_1t^a$ and (\ref{34}) into (\ref{14}):
\begin{equation}\label{35}
ah_1\left(\frac{R_h(0)^2}{h_0}-16bAM(0)^3\right)t^{a-1}+O(t^a)\geq
0.
\end{equation}
Note that $(a-1)$ is an odd integer, therefore if
$R_h(0)^2/h_0-16bAM(0)^3\geq (\leq)0$, GSL is not respected in
quintessence (phantom) phase before (after) the transition. Indeed
the black hole mass $M(0)$, gives the possibility that GSL becomes
respected in the quintessence era before the transition.

In the same way, for $\dot{R_h}(0)\neq 0$ we obtain
\begin{equation}\label{36}
bR_h(0)(h_0R_h(0)-1)+O(t)\geq 0.
\end{equation}
Therefore the generalized second law is respected at least in both
sides of the transition time provided that $\dot{R_h}(0)> 0$, in
agreement with the discussion after eq.(\ref{14}). Then the
continuity of $\dot{R_h}$, for $t$'s belonging to an open set
including $t=0$, results in $\dot{R_h}(t)> 0$.

In \cite{dav2}, it was shown that the future event horizon in the
quintessence model is a nondecreasing function of time. Using the
same method, in \cite{moh} it was proved that the future event
horizon is non increasing in phantom dominated era. In the first
view, combining these results leads to $\dot{R_h}(t=0)= 0$ which
prompts us to choose (\ref{34}). But this causes a conflict: near
the transition time, (\ref{34}) results in
$\dot{R_h}(t)=R_h(0)h_1t^a$, which is positive because $h_1>0$ and
$a$ is even, and this is in contradiction with the assumption
$\dot{R_h}(t>0)\leq 0$ proved in \cite{moh}. On the other hand if
we adopt $\dot{R_h}(t=0)\neq 0$, due to continuity of $\dot{R_h}$
(see (\ref{17})), there is an open set containing the transition
time in which the sign of $\dot{R_h}(t)$ is the same as the sign
of $\dot{R_h}(t=0)$, i.e. we have either $\dot{R_h}(t)<0$ in the
quintessence phase before the transition or $\dot{R_h}(t)>0$ in
the phantom phase after the transition. This conflict can be
solved by noting that the verifications of nondecreasing (non
increasing) behavior of $R_h$ in \cite{dav2}(\cite{moh}), were
based on the fact that the system remains in quintessence
(phantom) phase for all future time. So in the presence of
quintessence(phantom) to phantom (quintessence) phase transition,
it may be in general possible to have $\dot{R_h}(t)<0 (>0)$ for
some $t's$ in quintessence (phantom) era.

Following the above discussion we conclude {\it in an universe
which remains in phantom phase after the transition, GSL is not
respected in the neighborhood of transition time}, indeed for this
universe $\dot{R_h}\neq 0$ and in the vicinity of transition time
we have $\dot{R_h}<0$. To find an example of this situation see
\cite{moh}.

\end{document}